\documentclass[a4paper, aps, superscriptaddress, floatfix, twocolumn, longbibliography]{revtex4-1}

\usepackage{amsmath,amssymb}
\usepackage{graphicx}
\usepackage{dcolumn}
\usepackage{bm}
\usepackage[utf8]{inputenc}
\usepackage[caption=false]{subfig}
\usepackage{floatrow}
\usepackage{float}
\usepackage[rgb]{xcolor}
\usepackage{epsfig}
\usepackage{comment}
\usepackage{multirow}

\newcommand{\SM}[1]{SM}

\begin{document}

\title{Embedding-aided network dismantling}

\author{Saeed Osat} 
\affiliation{Max Planck Institute for Dynamics and Self-Organization (MPIDS), 37077 G{\"o}ttingen, Germany}

\author{Fragkiskos Papadopoulos}
\affiliation{Department of Electrical Engineering, Computer Engineering and Informatics, Cyprus University of Technology, 33 Saripolou Street, 3036 Limassol, Cyprus}

\author{Andreia Sofia Teixeira}
\affiliation{LASIGE,
Departamento de Inform\'atica, Faculdade de Ci\^encias, Universidade de Lisboa, 1749-016 Lisboa, Portugal}

\author{Filippo Radicchi}
\affiliation{Center for Complex Networks and Systems Research, Luddy School
  of Informatics, Computing, and Engineering, Indiana University, Bloomington,
  Indiana 47408, USA}


\begin{abstract}
Optimal percolation concerns
the identification of the
minimum-cost strategy for the destruction of
any extensive connected components in a network.
Solutions of such a dismantling problem
are important for the design of optimal strategies of disease containment based either on immunization or social distancing. 
Depending on the specific variant of the problem considered, 
network dismantling is performed via the removal of 
nodes or edges, and different cost functions are associated to the
removal of these microscopic elements.
In this paper, we show that network representations in geometric space can be used to solve several variants of the network dismantling problem in a coherent fashion. Once a network is embedded, 
dismantling is implemented using intuitive geometric strategies. We demonstrate that the approach well suits both Euclidean and hyperbolic network embeddings. Our systematic analysis on synthetic and real 
networks demonstrates that the performance of embedding-aided techniques is comparable to, if not better than, 
the one of the best dismantling algorithms currently available on the market.
\end{abstract}
                           
\maketitle



\section{Introduction}

Percolation theory aims at describing how the macroscopic connectedness of a network is affected by the removal of some of its microscopic elements~\cite{aharony2003introduction}. 
Percolation is among the most studied topics in statistical physics, especially for its relevance in the study of properties of materials, e.g., conductivity and porosity~\cite{kirkpatrick1973percolation}.
Since the advent of network science, the number of applications of percolation theory to real-world problems has constantly grown, and the literature on the topic has literally exploded~\cite{albert2002statistical, li2021percolation}.

In network science, 
the primary application of percolation theory 
is the study of the robustness of networks. 
The rationale is quite intuitive.
Being part of the same connected component is a
necessary condition for two nodes to
interact, thus large-scale connectedness represents a proxy for 
overall network function~\cite{cohen2001breakdown}.
Percolation allows the quantification of
the extent of damage that a network 
can tolerate before it is no longer able to guarantee such a condition. 
Percolation theory is useful 
not only to establish network robustness, but also in other contexts~\cite{li2021percolation}.
For example, the long-term behavior of some epidemic processes
is well predicted using the percolation framework~\cite{grassberger1983critical, pastor2015epidemic}, and
strategies for disease containment can be mapped to percolation
problems~\cite{cohen2003efficient}.

Percolation models assume the presence of an underlying network  where either nodes (site percolation) or edges (bond percolation) are removed according to some prescribed protocol~\cite{dorogovtsev2008critical}.
Nearest-neighbor non-deleted elements form connected components or clusters. The size of the clusters determines the regime of the network: (i) if only non-extensive clusters are present, then the network is in the non-percolating regime; (ii) if a giant connected component (GCC) spans a finite fraction of the network, then the system is in the percolating regime. 

Different deletion protocols may be considered, each defining a different percolation model with relevance for a specific problem
at hand. In the classical or ordinary model, individual elements are deleted randomly with uniform probability ~\cite{cohen2000resilience, newman2001random, dorogovtsev2008critical}. Real, heterogeneous networks display great robustness under this deletion protocol, as most of their elements should be removed before large-scale connectedness is lost. In targeted attacks,  the protocol prescribes elements to be removed on the basis of network centrality metrics~\cite{Albert2000, CallawayPRL, Cohen2000PRL}. In the context of site percolation, the model shows that heterogeneous networks, whose connectedness heavily relies on hubs, can be quickly dismantled by the removal of a small portion of their most central nodes.

The spirit of the model for targeted attacks is extremized in the so-called optimal percolation problem which consists in determining the minimum-cost deletion protocol able to bring the network into the non-percolating regime~\cite{Morone2015}. 
The problem was originally formulated for site percolation with unit cost of removal, and later generalized to bond percolation~\cite{ren2018underestimated} and to arbitrary cost functions associated to the removal of microscopic elements~\cite{GNetDismantling}. Finding the exact solution to the optimization problem requires testing all possible bipartitions of the microscopic elements of the network in two different sets of structural and non-structural elements. Structural elements are those that, if removed from the network, should fragment the system in non-extensive components. The number of possible bipartitions grows exponentially with the network size, thus the optimization problem is exactly solvable for very small systems only. Good approximate solutions can be achieved via simulated annealing (SA) optimization~\cite{NetDismantling}. However, the SA algorithm is not scalable. Existing algorithms able to approximate the solution of the problem in an efficient and effective way are based on rather different strategies. Many methods make use of a generic procedure where structural sets are constructed sequentially by adding one element at a time, and those elements are chosen on the basis of some ad-hoc network metric that is updated during the construction of the structural set. Methods of this class are based on collective influence~\cite{Morone2015, yuan2021fragmenting}, betweenness centrality~\cite{wandelt2018comparative}, non-backtracking centrality~\cite{NetDismantling}, explosive immunization~\cite{EI}, COREHD~\cite{COREHD}, and articulation points~\cite{tian2017articulation}, just to mention a few of them.
Another class of recent approaches takes advantage of machine learning methods to perform dismantling~\cite{fan2020finding, grassia2021machine}. Machines are trained on a huge number of small synthetic networks where the ground-truth solution of the dismantling problem can be obtained via brute-force search; these machines are then used efficiently and effectively to dismantle large-scale real networks. 
Finally, some methods existing on the market rely on graph embedding.
In Refs.~\cite{GNetDismantling, ren2018underestimated} for example, nodes are mapped into a one-dimensional space where their coordinates are given by the components of the first non-trivial eigenvector of specifically designed Laplacian operators. In Ref.~\cite{wandelt2020community} instead, the map is determined by the community structure of the network, so that nodes are embedded in a space that is not metric. Once the network is embedded in space, then a deletion protocol based on the map is used to construct a solution of the optimal percolation problem.  

Many of the above algorithms focus on the simplest formulation of the problem where dismantling is performed by removing nodes, and the cost of removal is equal to the size of the structural set. 
Other variants of the problem are considered sporadically. 
For example, Ref.~\cite{ren2018underestimated} studies the bond-percolation version of the problem. Other important variants of the problem are those considered by Bellingeri {\it et al.} who study optimal site percolation on weighted networks~\cite{bellingeri2020comparative}, and by Lokhov {\it et al.} who focus on optimal strategies of immunization for spreading processes~\cite{lokhov2017optimal}.

 In this paper, 
 we leverage embedding of networks in 
 geometric
 space to perform efficient network dismantling. We show that the same type of methodology can be used for both Euclidean and hyperbolic embeddings. Furthermore, we demonstrate that the same embedding can be fruitfully used to provide effective solutions to various variants of the optimal percolation problem based on the removal of nodes or edges, and constrained by different cost functions. We systematically apply the proposed methods on a 
 corpus of $50$ real-world networks.
 We find that the performance of embedding-aided dismantling algorithms is comparable to the one of the best methods existing on the market.
 Further, we apply the methods to synthetic graphs generated according to the $\mathbb{H}^2$ model
~\cite{aldecoa2015hyperbolic} and the Lancichinetti-Fortunato-Radicchi  model~\cite{lancichinetti2008benchmark}. 
Both these models generate networks that are embedded in an underlying space; moreover, they are characterized by parameters that allow to tune the strength of the relationship between network structure and imposed embedding. We find that the proposed embedding-aided methods outperform the other dismantling algorithms only when topology and embedding are strongly correlated. Performances of the various methods become comparable when such a relationship is weak.


\section{Results}

\subsection{Geometric approach to network dismantling}

The problem we consider in this paper is the identification of the minimum-cost strategy for the destruction of any extensive connected component in unweighted and undirected networks, see Methods section for a formal definition. The destruction is performed by the removal of microscopic elements, either nodes (site percolation) or edges (bond percolation). The optimization problem is constrained by the cost function $F(\mathcal{S})$, which quantifies the cost associated with the removal of the elements of an arbitrary set $\mathcal{S}$. We consider the unit-cost function for both bond and site percolation, and the degree-cost function for the site-percolation problem only. 

Exact solutions to the above problem are not feasible due to the exponentially growing number of possible sets that must be considered as possible solutions to the problem. Approximate solutions are obtained via so-called dismantling algorithms. 
 Assuming that there are $T$ total microscopic elements that can be removed from the network, the output of a dismantling algorithm is the sequence of sets $\tilde{\mathcal{S}}_0, \tilde{\mathcal{S}}_1, \ldots, \tilde{\mathcal{S}}_T$, with $\tilde{\mathcal{S}}_{t-1} \subset \tilde{\mathcal{S}}_{t}$ for all $t=1, \ldots T$.
 The sequence of sets $\tilde{\mathcal{S}}_{t}$ indicates how to dismantle the network. Clearly, this sequence represents only an approximation of the
 ground-truth solution of the dismantling problem.  

In this paper, we introduce a family of dismantling algorithms based on network embedding. The input network is first embedded in geometric space, meaning that each node $i$ of the graph is mapped to a point $\vec{v}_i$ in an underlying $d$-dimensional vector space. The map is used to iteratively create network bipartitions, and the sets $\tilde{\mathcal{S}}_{t}$ are constructed by adding blocks of inter-cluster elements identified at each iteration, see Methods for details. 

The above geometric recipe has been introduced by Ren {\it et al.} and 
involves the embedding of the graph in one-dimensional space using graph Laplacian operators~\cite{GNetDismantling, ren2018underestimated}. 
Specifically, the sign of the components of the eigenvector associated to the second smallest eigenvalue of Laplacian-like operators are used to create bipartitions. Essentially, the dismantling of the network is approached by solving small-scale minimum-cut problems.

The same idea can be generalized to any type of embedding that  captures structural similarity among nodes in the graph. Community structure is a way of performing such a task with a non-metric embedding; community structure has been exploited in the context of network dismantling in Ref.~\cite{wandelt2020community}. In this paper, we consider embeddings in vector spaces, either hyperbolic or Euclidean spaces. We employ two popular embedding methods, namely Mercator for hyperbolic embedding~\cite{Mercator_repo, Garc_a_P_rez_2019} and Node2vec for Euclidean embedding~\cite{grover2016node2vec}.
The motivation for going beyond already existing embedding-aided dismantling algorithms  is two-fold. First, we believe that the high-dimensionality of the embedding space should allow us to capture additional features compared to the one-dimensional Laplacian embedding. Second, we believe that the 
geometric
nature of the embedding space should allow us to obtain a more nuanced definition of clusters compared to community-structure embeddings.

\begin{figure}[!htb]
\captionsetup{farskip=0pt}
\includegraphics[width=0.9\columnwidth]{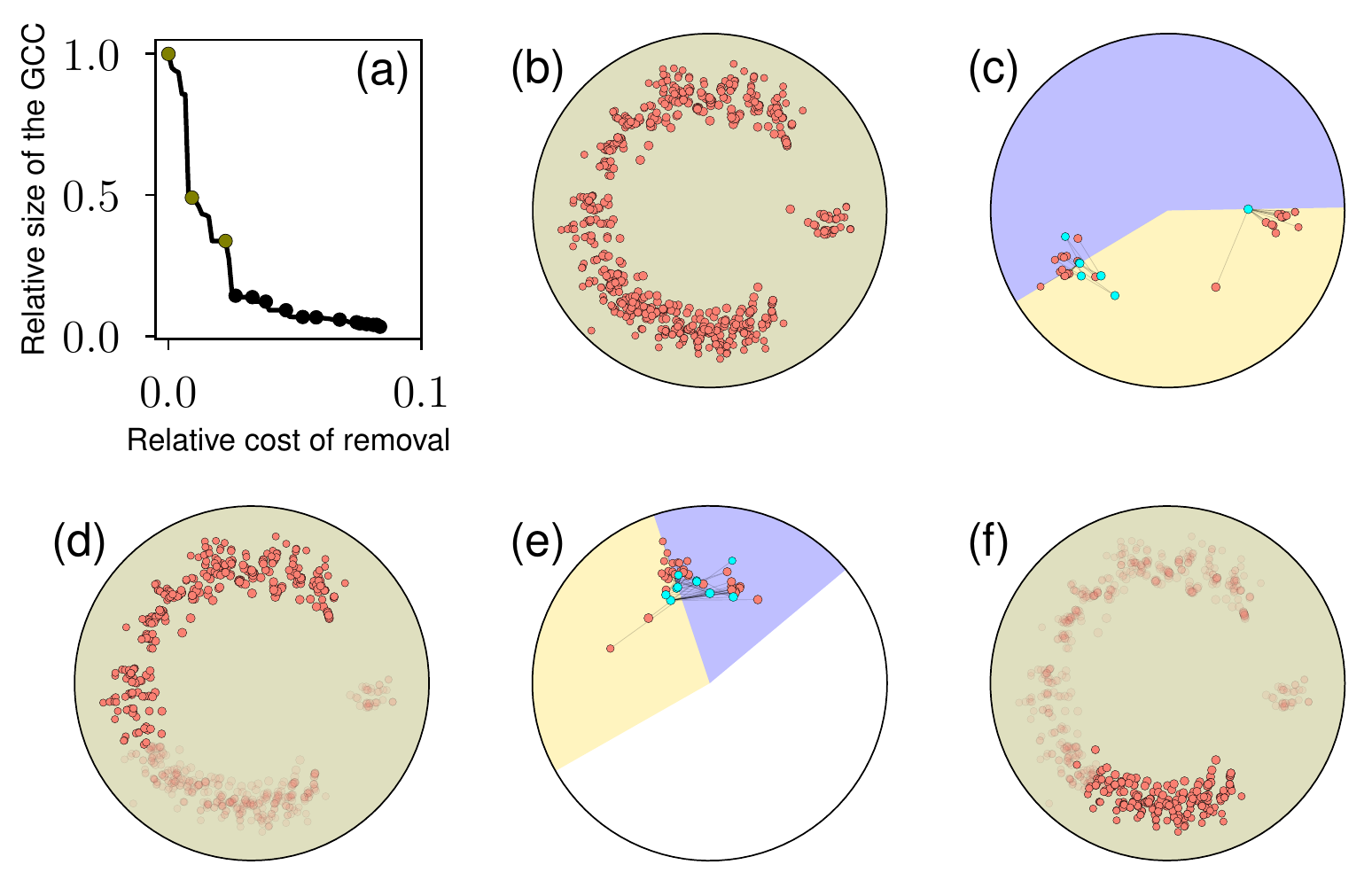}
\caption{{\bf Network dismantling aided by hyperbolic embedding.}
(a) Relative size of the giant connected component (GCC) as a
function of the relative cost of removal. We consider the case of unit-cost site percolation, thus the x-axis values represent the fraction of nodes removed in the network. 
Nodes are removed according to the  dismantling algorithm that leverages hyperbolic embedding. The network under consideration is the network connecting persons who committed the same crimes~\cite{KONECT}. A point on the curve denotes an iteration of the algorithm; the curve between two points is obtained by randomly sorting nodes deleted at a given iteration (see Methods for details). The insets illustrate the basic mechanisms of the algorithm and refer to the first three stages of the algorithm. (b) The network is first embedded in hyperbolic space.  (c) The hyperbolic disk is sliced in two parts, cyan nodes are identified as the nodes to be inserted in the structural set. (d-f) In the following stages, the above steps are iterated. Specifically, the coordinates of the nodes of the first cluster (d) are used to find other nodes to be inserted in the structural set (e). The same procedure is applied independently to the nodes belonging to the other cluster identified at the first stage of the algorithm (f).}
\label{fig1}
\end{figure}


In Fig.~\ref{fig1}a, we show an example of the application of the hyperbolic-embedding-aided dismantling algorithm to solve the optimal site-percolation problem with unit cost on the crime network obtained from the projection of the bipartite network of crimes and individuals~\cite{KONECT}. A technical description of the algorithm is provided in the Methods section. Here, we just describe it in simple terms to give an intuition of how the method works.
First, we embed the network in hyperbolic space, as shown in Fig.~\ref{fig1}b. Then, we split the network in two clusters by slicing the hyperbolic disk in two parts. Each slice of the disk contains the same number of nodes. We note that there are multiple ways to slice the disk, and some of them lead to better solutions to the optimal percolation problem than others. However, we do not observe huge variations in performance depending on how the two slices of the disk are obtained (see Figs.~S1 and~S2). The actual separation in clusters of the two slices is achieved by removing the smallest number of nodes that lead to such a separation, highlighted in cyan in panel Fig.~\ref{fig1}c. Those nodes are added in random order to the structural set to reduce the size of the GCC, see Fig.~\ref{fig1}a. We then apply the same operation to each of the resulting clusters, see Figs.~\ref{fig1} d-f. We do not need to re-embed the clusters, rather we can simply re-use the known coordinates of the remaining nodes to cut in half the corresponding slices in the hyperbolic disk, and then remove the minimal number of nodes to split each slice in two disconnected clusters. The entire procedure is iterated over and over, until the network is fully dismantled. 

The same exact principle can be easily extended to deal with a different embedding. For Euclidean embeddings such as those created by Node2vec for example, the $k$-means algorithm~\cite{lloyd1982least} with $k=2$ is used to determine the bipartitions required by the dismantling protocol, see Methods for details.

As described in the Methods section,
dismantling the network requires a time that grows slightly more than linearly with the network size.
The dismantling recipe, however, assumes that the embedding of the network is given, but such an operation may require a number of computations that grows super-linearly with the network size, thus dominating the actual time complexity of the entire dismantling procedure.
This is the case of the hyperbolic-embedding-aided dismantling method where the embedding algorithm requires a time that scales as the square of the network size. Node2vec instead requires a time that grows linearly with the system size, thus the resulting dismantling algorithm scales quasi-linearly with the network size. 

The above geometric method is easily adapted to any variant of the network dismantling problem. For example, the bond-percolation version is  obtained by splitting clusters via link removal instead of node removal. Details of the various algorithms are provided in the Methods section.

\subsection{Performance of geometric dismantling}

\begin{figure*}[!htb]
\captionsetup{farskip=0pt}
\includegraphics[width=0.9\columnwidth]{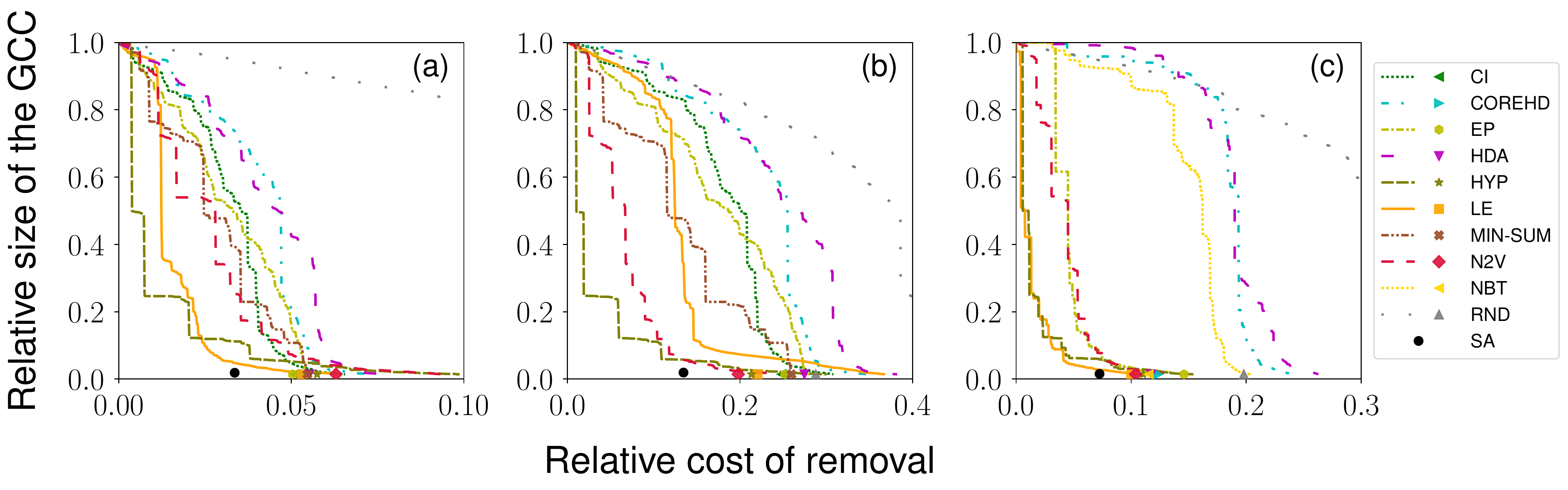}
\caption{{\bf Optimal percolation on the US power grid.}
(a) Relative size of the giant connected component (GCC) as a function of the fraction of removed nodes. Different curves correspond to solutions obtained via the various dismantling methods. Symbols present solutions to the problem of Eq.~(\ref{eq:opt_perc2}) obtained by the greedy post-processing strategy started from the solution of a given algorithm. The network considered here is the US power grid~\cite{watts1998collective}. (b) Same as in panel (a), but for the  optimal site-percolation problem with degree cost. (c) Same as in panel (a), but for  optimal bond percolation with unit cost.
}
\label{fig2}
\end{figure*}

\begin{figure*}[!htb]
\captionsetup{farskip=0pt}
\includegraphics[width=0.9\columnwidth]{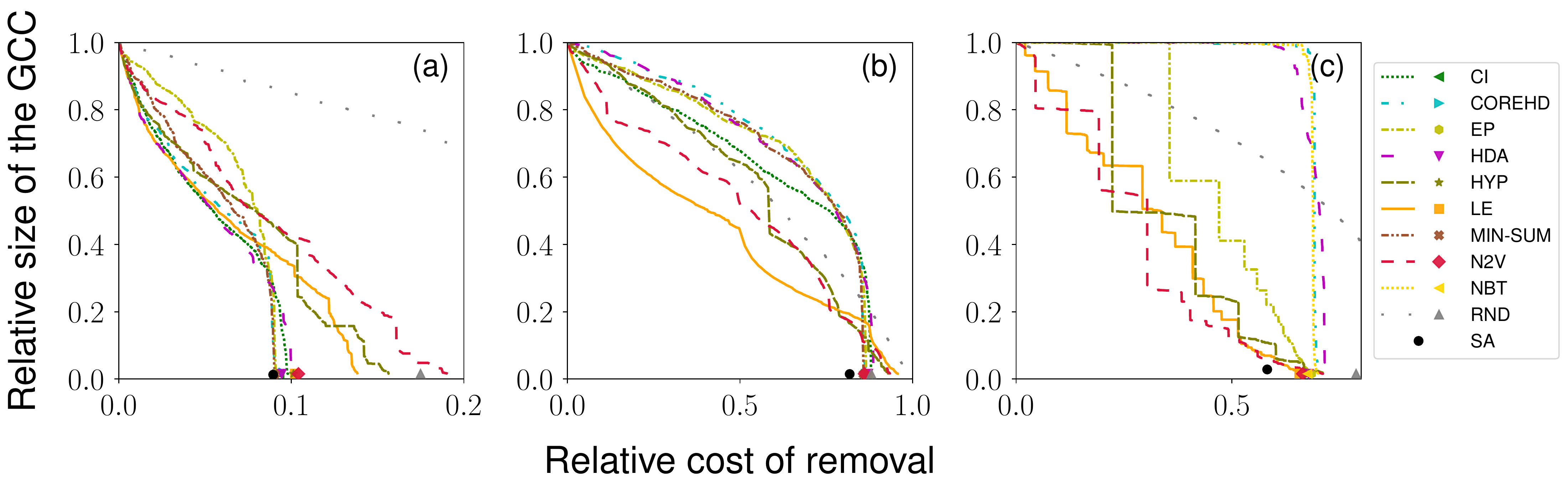}
\caption{{\bf Optimal percolation on the Proteome network.}
Same as in Fig.~\ref{fig2}, but for the Proteome network~\cite{net_proteome}.}
\label{fig3}
\end{figure*}

We compare the performance of our proposed algorithms against those of well-established baselines and top-performing algorithms existing in the market, see Methods for details. A specific example is displayed in Fig.~\ref{fig2}. There, various dismantling algorithms are applied to the network representing the topology of the US power grid~\cite{watts1998collective}. In the figure, we display how the relative size of the GCC decreases as a function of the cost associated to the removal of microscopic elements from the network. The quicker the decrease is, the better the approximate solution of the specific algorithm at hand is. The area under the percolation curve is named as robustness and is generally used as a metric of performance for dismantling algorithms~\cite{schneider2011mitigation}, see Eq.~(\ref{eq:evaluation}). We denote the robustness metric as $R$. As expected, all algorithms produce approximate solutions that are better than those obtained via random removal (RND). The hyperbolic-embedding-aided dismantling (HYP) outperforms the other algorithms in the three variants of the dismantling problem; the least-performing algorithm is the one based on adaptive degree centrality (HDA). 

As an additional metric of performance, we also display the cost function of the structural set required to decrease the size of the GCC below the square root of the network size, see Eq.~(\ref{eq:opt_perc2}). 
This specific threshold value is just a convention used to determine whether all connected components are not extensive~\cite{EI}. We refer to this metric as the  dismantling cost of the network, and denote it as $q_c$.

Solutions of the various algorithms can be further improved by a greedy post-processing technique, see Methods for details. The technique was introduced in Ref.~\cite{NetDismantling} for the case of unit-cost site percolation. Here, we generalize it to the various variants of the optimal percolation problem. The technique basically consists in removing from the structural set all unnecessary elements, i.e., those elements that if removed from the set do not lead to the emergence of an extensive GCC.  In Fig.~\ref{fig2}, we display solutions that have been improved with this technique as single points denoting the value of the dismantling cost that is reached after greedy optimization. All solutions become similar after being greedily optimized, displaying performance that is bounded by RND and simulated annealing (SA) optimization. Please note that the greedy post-processing just minimizes the dismantling cost of the network. The technique is not designed to speedup the actual dismantling, thus it does not necessarily reduce the value of the robustness metric.

The curves displayed in 
Fig.~\ref{fig2} indicate that, before the application of the greedy optimization step, there are variations in performance depending on the specific algorithm and the specific variant of the percolation problem considered.  Variability in performance also depends on the specific type of network considered. In Fig.~\ref{fig3}, we repeat the same analysis as in Fig.~\ref{fig2}, but on the Proteome network~\cite{net_proteome}. An apparent change in relative performance among the various methods is visible. For example, the Node2vec-embedding-aided method (N2V) is the least performing method in the site-percolation problem with unit cost  (Fig.~\ref{fig3}a), while it was among the best in dismantling the US power grid network  (Fig.~\ref{fig2}a).

\subsection{Systematic analysis of real-world networks}


\begin{figure}[!htb]
\captionsetup{farskip=0pt}
\includegraphics[width=1\columnwidth]{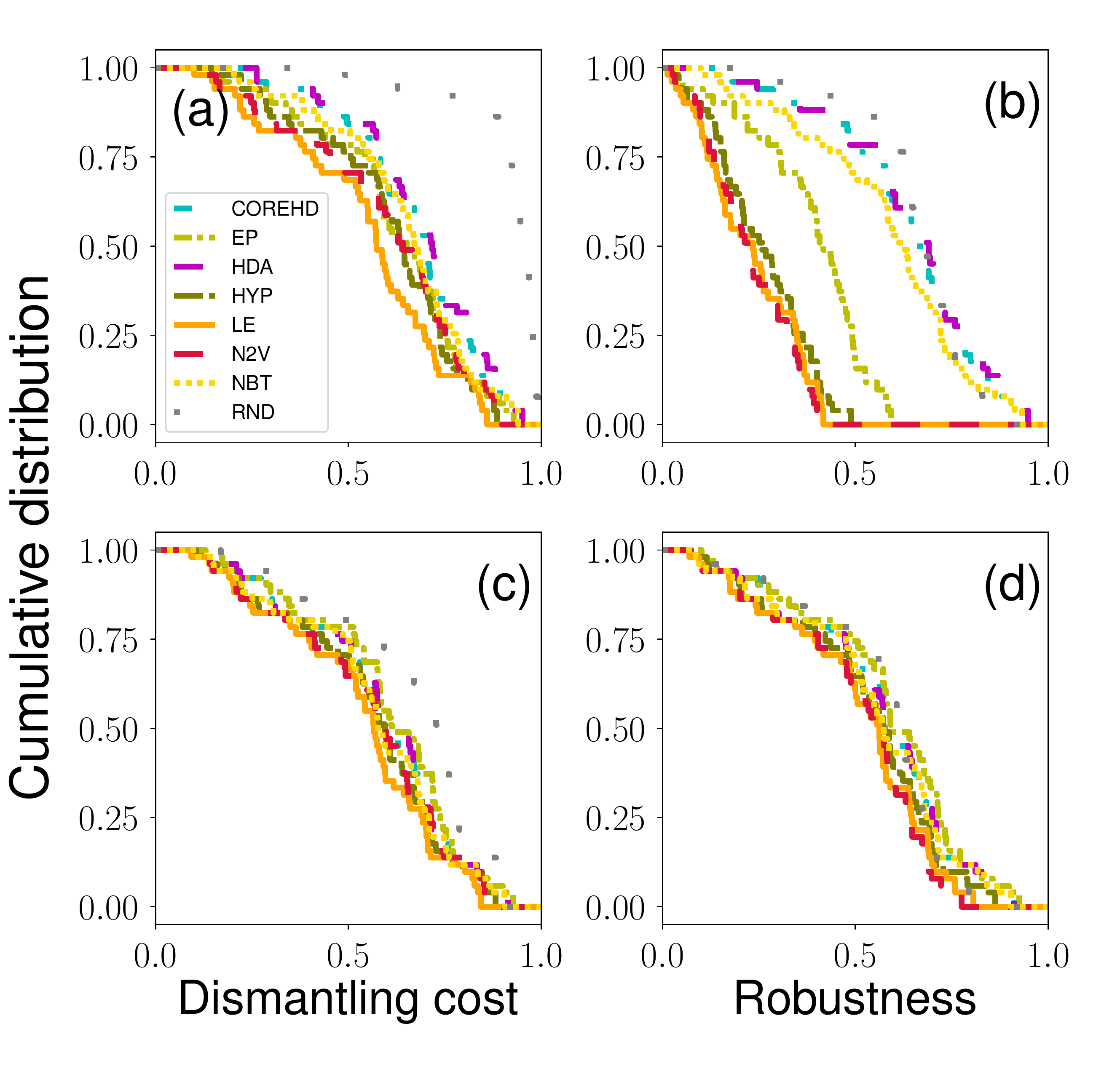}
\caption{
{\bf Optimal bond percolation in real-world networks.}
(a) Cumulative distribution function of the relative dismantling cost of the various algorithms in the solution of the optimal bond percolation with unit cost. The distribution is evaluated on a corpus of 50 real-world networks. The dismantling cost is defined in Eq.~(\ref{eq:evaluation_a}). (b) Same as in (a), but for the robustness metric, see Eq.~(\ref{eq:evaluation}). (c and d) Same as in (a) and (b), respectively, but for solutions obtained after the application of the greedy post-processing technique.
} 
\label{fig4}
\end{figure}

\begin{table*}[!htb]
\resizebox{0.8\textwidth}{!}{%
\begin{tabular}{l c cccc c cccc c cccc}
 & & \multicolumn{4}{c}{Bond} &  & \multicolumn{4}{c}{Site (unit cost)} & & \multicolumn{4}{c}{Site (degree cost)} \\
\cline{3-6} \cline{8-11} \cline{13-16}
& & \multicolumn{2}{c}{Regular} & \multicolumn{2}{c}{Greedy} &
& \multicolumn{2}{c}{Regular} & \multicolumn{2}{c}{Greedy} &
& \multicolumn{2}{c}{Regular} & \multicolumn{2}{c}{Greedy}
\\
Method & & $q_c$ & $R$ &  $q_c$ & $R$ & & $q_c$ & $R$ & $q_c$ & $R$ & & $q_c$ & $R$ & $q_c$ & $R$
\\
\cline{3-6} \cline{8-11} \cline{13-16}
\cline{1-1}
\cline{3-6} \cline{8-11} \cline{13-16}
 CI  & & - & - & - & -  &
 & $0.18$ & $0.12$ & $0.14$ & $0.9$  &
 & $0.80$ & $0.55$ & $0.75$ & $0.51$ 
\\
\cline{1-1}
\cline{3-6} \cline{8-11} \cline{13-16}
COREHD  & & $0.67$ & $0.63$ & $0.57$ & $0.55$  &
& $0.15$ & $0.10$ & $\boldsymbol{0.13}$ & $\boldsymbol{0.09}$ &
& $0.78$ & $0.61$ & $0.74$ & $0.54$  
\\
\cline{1-1}
\cline{3-6} \cline{8-11} \cline{13-16}
 EP  & & $0.61$ & $0.38$ & $0.60$ & $0.58$   &
 & $\boldsymbol{0.14}$ & $0.10$ & $\boldsymbol{0.13}$ & $0.09$   &
 & $\boldsymbol{0.76}$ & $0.56$ & $0.74$ & $0.55$ 
 \\
\cline{1-1}
\cline{3-6} \cline{8-11} \cline{13-16}
 HDA  & & $0.69$ & $0.65$ & $0.57$ & $0.55$  &
 & $0.15$ & $0.09$ & $0.13$ & $\boldsymbol{0.09}$   &
 & $0.80$ & $0.61$ & $0.75$ & $0.54$ 
\\
\cline{1-1}
\cline{3-6} \cline{8-11} \cline{13-16}
 HYP  & & $0.60$ & $0.25$ & $0.55$ & $0.52$ &
 & $0.19$ & $0.09$ & $0.13$ & $0.09$  &
 & $0.79$ & $0.43$ & $\boldsymbol{0.71}$ & $0.49$ 
 \\
\cline{1-1}
\cline{3-6} \cline{8-11} \cline{13-16}
 LE  & & $\boldsymbol{0.54}$ & $\boldsymbol{0.22}$ & $\boldsymbol{0.53}$ & $\boldsymbol{0.50}$ &
 & $0.16$ & $\boldsymbol{0.09}$ & $0.13$ & $0.09$ &
 & $0.83$ & $\boldsymbol{0.36}$ & $0.72$ & $\boldsymbol{0.46}$  
 \\
\cline{1-1}
\cline{3-6} \cline{8-11} \cline{13-16}
 MIN-SUM  & & - & - & - & - &
 & $\boldsymbol{0.13}$ & $0.10$ & $0.13$ & $0.09$   &
 & $\boldsymbol{0.74}$ & $0.58$ & $0.74$ & $0.55$ 
\\
\cline{1-1}
\cline{3-6} \cline{8-11} \cline{13-16}
 N2V  & & $\boldsymbol{0.59}$ & $\boldsymbol{0.22}$ & $\boldsymbol{0.55}$ & $\boldsymbol{0.50}$ &
 & $0.20$ & $\boldsymbol{0.09}$ & $0.14$ & $0.09$ &
 & $0.78$ & $\boldsymbol{0.40}$ & $\boldsymbol{0.71}$ & $\boldsymbol{0.48}$
\\
\cline{1-1}
\cline{3-6} \cline{8-11} \cline{13-16}
 NBT  & & $0.64$ & $0.58$ & $0.56$ & $0.54$  &
 & - & - & - & - &
 & - & - & - & - 
\\
\cline{1-1}
\cline{3-6} \cline{8-11} \cline{13-16}
 RND  & & $0.92$ & $0.66$ & $0.66$ & $0.56$  &
 & $0.59$ & $0.31$ & $0.17$ & $0.11$  &
 & $0.92$ & $0.54$ & $0.74$ & $0.50$ 
 \end{tabular}
 }
 \caption{{\bf Optimal percolation in real-world networks.} For each dismantling method, we report the average value across the corpus of 50 real-world networks of the  dismantling cost $q_c$ [Eq.~(\ref{eq:evaluation_a})] and the robustness metric $R$ [Eq.~(\ref{eq:evaluation})]. We separate results depending on the specific variant of the optimal percolation problem considered. Also, we report results valid before and after the application of the greedy post-processing strategy. Data for optimal bond percolation, optimal site percolation with unit cost, and optimal site percolation with the degree cost are the same as in Figs.~\ref{fig4}, S3, and S4, respectively. Performance values of the top two performing methods for each category are highlighted with bold fonts. Visualized values are rounded to two significant digits, but comparisons are performed before rounding.}
 \label{table}
\end{table*}


We perform a systematic analysis on a corpus of $50$ real-world networks. For each network, we consider the three variants of the optimal percolation problem ( i.e., bond percolation with unit cost, site percolation with unit cost, and site percolation with degree cost), we apply each of the 
dismantling algorithms considered in this analysis, and measure the performance in terms of the relative dismantling cost $q_c$ of  Eq.~(\ref{eq:evaluation_a}) and the robustness metric $R$ of Eq.~(\ref{eq:evaluation}).  Detailed results are reported in SM, and summarized in Figs.~\ref{fig4}, S3 and S4, and in Table~\ref{table}.
Embedding-aided algorithms display performance comparable to the one of the other well-established methods for network dismantling in all variants of the problem.
Notably, the methods based on Laplacian Embeddings (LE) and N2V excel in all tasks. 

The solution of each method is refined using the greedy post-processing strategy introduced in Ref.~\cite{NetDismantling}. 
Whereas the ranking of the various methods based on performance is not much affected by the post-processing technique, the gap in performance between the various methods is narrowed. 
Essentially, greedy post-processing leads to almost equivalent solutions irrespective of the starting structural set generated by a given method. The only clear exception is RND, which still displays a clear gap with respect to the other methods in spite of the application of the greedy post-processing step. Also, we remark that the greedy post-processing technique always reduces the dismantling cost of the set of structural elements identified by an algorithm. However, such an improvement in the metric is generally accompanied by a loss of performance in terms of robustness, see Fig.~S5 for example. The effect is systematic in all variants of the percolation problem, except for site percolation with unit cost.


\subsection{Systematic analysis of synthetic networks}

We conclude our analysis by studying the performance of the various dismantling methods on synthetic networks generated according to
the hyperbolic $\mathbb{H}^2$ model
~\cite{aldecoa2015hyperbolic} and the Lancichinetti-Fortunato-Radicchi (LFR) model~\cite{lancichinetti2008benchmark}.
The use of these models is motivated by their ability to reproduce topological properties that resemble the ones observed in real-world networks, as for example heterogeneous degree distribution, high clustering, and modular structure. 
Also, these models generate networks that are naturally embedded in some underlying space, either geometric or not, thus allowing us to verify how important this property is for the actual performance of the various dismantling methods. 

\begin{figure*}
\captionsetup{farskip=0pt}
\includegraphics[width=0.9\textwidth]{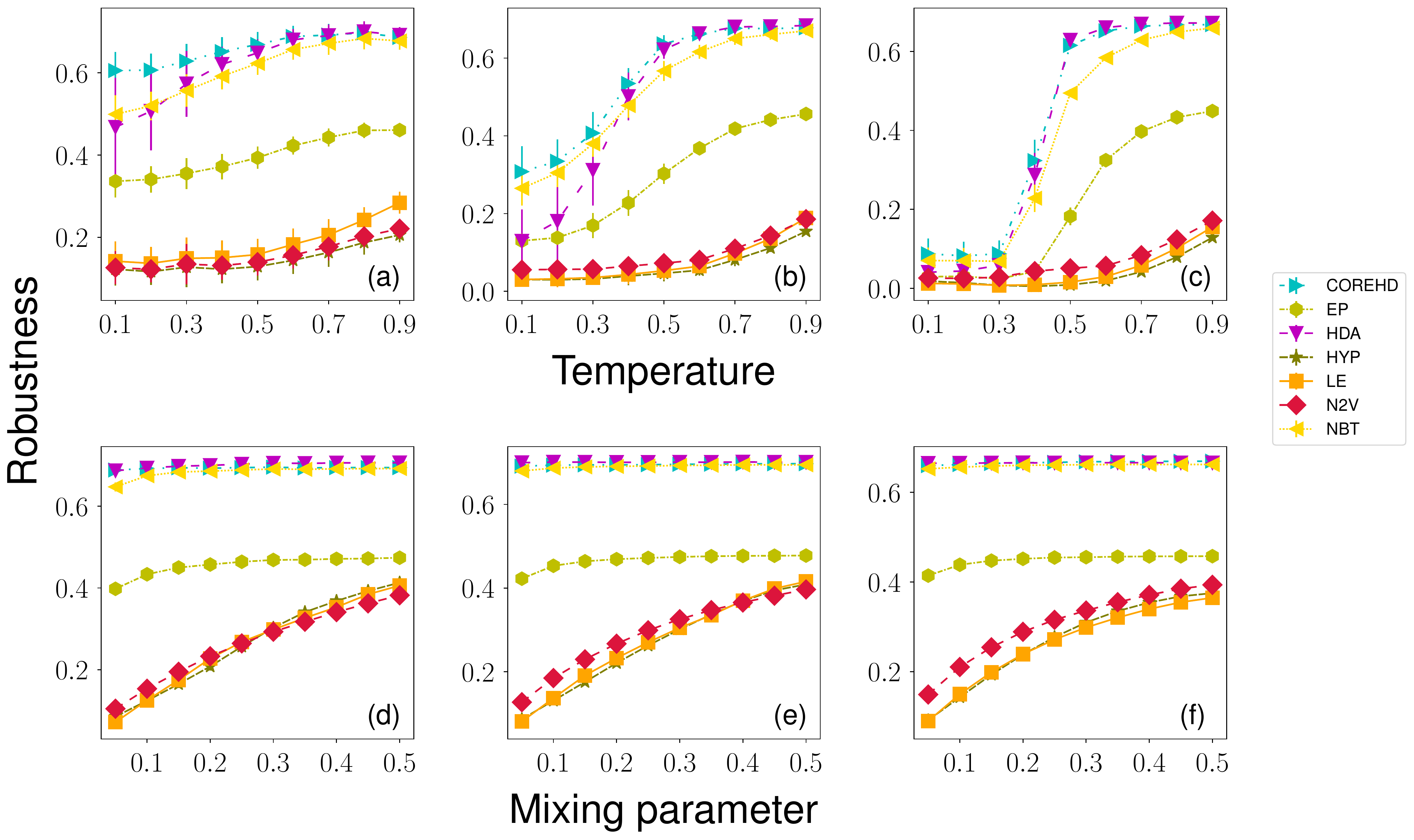}
\caption{{\bf Optimal bond percolation in synthetic networks.}
(a) We apply the various dismantling methods to instances of the $\mathbb{H}^2$ model (see Methods for details). 
The degree distribution is a power law with exponent $\gamma = 2.2$. We construct networks for different values of the temperature parameter of the model, and measure 
the robustness metric defined in Eq.~(\ref{eq:evaluation}).
Each point is an average over $100$ different instances of the $\mathbb{H}^2$ model. (b) Same as in (a), but for $\gamma = 2.6$. (c) Same as in (a), but for $\gamma = 3.5$. (d) We apply the various dismantling methods to instances of the Lancichinetti-Fortunato-Radicchi model (see Methods for details). The degree distribution is a power law with exponent $\gamma = 2.2$. We measure 
the robustness metric as a function of the mixing parameter of the LFR model. (e) Same as in (d), but for $\gamma = 2.6$. (f) Same as in (d), but for $\gamma = 3.5$.}
\label{fig5}
\end{figure*}

Some results are reported in Fig.~\ref{fig5}; full results are instead displayed in Figs.~S6-S11. For bond percolation with unit cost, the main outcome of our analysis is two-fold. First, the performance of all methods decreases as the
relationship between network topology and imposed embedding weakens. 
This fact can be clearly appreciated for the $\mathbb{H}^2$ model by monitoring how performance varies with the temperature parameter, and for the LFR model by monitoring how performance changes as a function of the mixing parameter.  Second, embedding-aided dismantling methods outperform the other methods. The gap in performance is particularly apparent for networks with homogeneous degree distributions. The above considerations are valid either if performance is measured in terms of dismantling cost or robustness. The gap in performance between centrality-based and embedding-aided methods is not filled even if  greedy post-processing is applied to the structural sets.

 For site percolation with unit or degree cost of removal, we find that centrality-based outperform embedded-aided methods. The gap in performance, however, disappears once the greedy post-processing technique is applied to the structural sets found by the various methods. 

We remark that the job of HYP on $\mathbb{H}^2$ networks is facilitated by the fact that no embedding is actually learned from the topology, rather ground-truth coordinates of the nodes in the hyperbolic space are used to dismantle the network.

\section{Conclusions}

The results of this paper clearly show that embedding a network in geometric space can be  used to design simple but effective algorithms to dismantle it. Such geometric dismantling techniques are rather general. They can be adapted to various types of embeddings, and they appear useful in solving different variants of the optimal percolation problem. The proposed techniques are also computationally efficient. Once the network is embedded, dismantling is performed in a time that grows slightly super-linearly with the network size. However, it is important to keep in mind that embedding a network may require a time that grows more than linearly with the system size. For example, embedding a network in the hyperbolic space generally requires a time that grows quadratically with the network size~\cite{Garc_a_P_rez_2019}; obtaining a map of the network in Euclidean space with Node2vec requires instead a time that grows linearly with the network size~\cite{grover2016node2vec}. 
The performance of embedding-aided dismantling methods is comparable to the one achieved by other methods existing in the market that are based on different heuristics. The general message is that embedding-aided algorithms excel in bond percolation, whereas they are outperformed by centrality-based methods in site percolation. Eventual gaps in performance between the various dismantling methods are anyway filled by applying the greedy post-processing technique originally proposed in Ref.~\cite{NetDismantling} for site percolation, and here generalized to the other variants of the optimal percolation problem. In essence, optimal performance can be achieved by first applying a sufficiently effective method to dismantle a network, and then reducing the cost of the structural set identified by the algorithm via greedy optimization. 

Due to the similarity in performance between the various algorithms, the use of a computationally efficient method such as Node2vec may be naively preferred over other methods to perform the embedding necessary to geometrically dismantle a network. We stress, however, that computational time is not the only important aspect to consider here. Hyperbolic maps consist of only two coordinates per node, making them particularly suited to provide meaning and intuitive visualizations. The same consideration does not apply to Euclidean embeddings which instead are generally performed for much larger values of the space dimension. Also, popular methods that perform Euclidean embedding often require the calibration of several parameters; this procedure is much less expensive, if not totally absent, in algorithms that embed networks in hyperbolic space.

\section{Methods}

\subsection{The optimal percolation problem}

We consider an undirected and unweighted network with $N$ nodes. Pairwise interactions among nodes are encoded in the symmetric adjacency matrix $A$. If an edge exists between nodes $i$ and $j$, then $A_{ij}=1$;  $A_{ij}=0$ otherwise. 

Large-scale connectedness of the network is quantified in terms of the fraction of nodes that belong to the giant connected component (GCC) of the network as $P_{\infty} = \frac{N_{GCC}}{N}$, where $N_{GCC}$ is the number of nodes in the GCC of the network.
We indicate with $\mathcal{T}$ the set of all microscopic elements, either nodes or edges, of the network; the size of the GCC can be reduced by removing from the network elements belonging to a subset $\mathcal{S} \subseteq \mathcal{T}$. We refer to the subset $\mathcal{S}$ as the structural subset of the network, and to the elements within the set $\mathcal{S}$ as the structural elements of the network.
Without loss of generality, we assume that, when all microscopic elements are present, 
the network is composed of a single connected component. In other words, if the set of structural elements is empty, i.e., $\mathcal{S} = \emptyset$, then $P_{\infty} (\emptyset) = 1$. The removal of microscopic elements from a non-empty set $\mathcal{S}$ from the network causes a reduction of the GCC, i.e., $P_{\infty}(\mathcal{S}) \leq 1$. Clearly, the removal of all microscopic elements leads to the smallest size of the GCC, i.e., $P_{\infty}(\mathcal{T}) = 0$ for site percolation and $P_{\infty}(\mathcal{T}) = 1/N$ for bond percolation.

Optimal percolation can be seen as the constrained minimization problem
\begin{equation}
    \mathcal{S}^* (C) = \arg \, \min_{\mathcal{S} | F(\mathcal{S}) = C} P_{\infty}(\mathcal{S}) \; .
    \label{eq:opt_perc}
\end{equation}
The constraint is imposed on the value of the cost function 
$F(\mathcal{S})$ of removing elements of the set $\mathcal{S}$
from the network.
In the original formulation of the problem by Morone and Makse, the cost associated to the set $\mathcal S$ is identical to the size of the set, i.e., $F(\mathcal S) = |\mathcal S|$~\cite{Morone2015}. However, an arbitrary cost can be associated to the set in the so-called generalized dismantling problem framed by Ren {\it et al.}~\cite{GNetDismantling}.
To be a meaningful cost function, we only require that $F(\mathcal{S}) \geq 0$ for all sets $\mathcal{S}$, and $F(\mathcal{S} \cup \{s\}) \geq F(\mathcal{S})$ for every set $\mathcal{S}$ and any microscopic network element $s$.
It also natural to assume that $F(\emptyset) = 0$.

In this paper, we consider three main formulations of the optimal percolation problem:
(i) unit-cost optimal site percolation,
(ii) optimal site percolation with degree cost, and 
(iii) unit-cost optimal bond percolation.
In formulations (i) and (ii), network dismantling is performed via the removal of  nodes; in (iii), network dismantling is performed via the removal of edges. In the unit-cost version of the problem, the cost function associated to the set $\mathcal{S}$ equals its size, i.e., $F(\mathcal{S}) = |\mathcal{S}|$. The degree-cost function of variant (ii) is defined as $F(\mathcal{S}) = \sum_{s \in \mathcal{S}} \, k_s - \sum_{s,t \in \mathcal{S}} A_{st}$, where $k_s$ is the degree of node $s$, the sums run over all nodes in the set  $\mathcal{S}$, and edges shared by nodes within the set  $\mathcal{S}$ are counted only once.

An important aspect in the characterization of the optimization problem is the identification of the minimum-cost set able to lead to the disappearance of a macroscopic GCC~\cite{Morone2015}. Such a condition is defined in the problem
\begin{equation}
    \mathcal{S}_c = \arg \min_{\mathcal{S} | P_\infty(\mathcal{S}) \leq 1/\sqrt{N}} F(\mathcal{S}) \; .
    \label{eq:opt_perc2}
\end{equation}
Essentially, only sets $\mathcal{S}$ that are able to reduce $P_\infty$ below the conventional threshold value $1/\sqrt{N}$ are considered as potential solutions to the problem~\cite{EI}. 

\subsection{Approximate solutions of the optimal percolation problem}

The optimal percolation problem of Eq.~(\ref{eq:opt_perc}) is NP hard~\cite{Morone2015}. For example, in the optimal site percolation with unit cost, 
the exact solution of the problem requires to test all possible ${N \choose |\mathcal{S}|}$ sets that can be composed by choosing $|\mathcal{S}|$ nodes out of the $N$ total nodes in the network.

Exact solutions of the optimization problem can be obtained only for extremely small networks. However, many algorithms able to approximate solutions to the optimal percolation problem have been proposed. Some of these algorithms are described below. Without loss of generality, we indicate with $\tilde{\mathcal{S}}^*$ an approximate solution to the problem of Eq.~(\ref{eq:opt_perc}) obtained by a generic algorithm. Similarly, we use the notation $\tilde{\mathcal{S}}_c$ to denote an approximate solution to the problem of Eq.~(\ref{eq:opt_perc2}). 

Many optimization algorithms construct approximate solutions to the optimal percolation problem sequentially, meaning that the set corresponding to the proposed solution is built by adding one element at time. We indicate
with $\tilde{\mathcal{S}}_t$ the approximate solution of a generic sequential algorithm when the set is composed of exactly $t$ elements, i.e., $|\tilde{\mathcal{S}}_t| = t$. If there are $T$ total microscopic elements in the network, i.e., $|\mathcal{T}| = T$, the sequential algorithm generates $T+1$ total sets, i.e., $\tilde{\mathcal{S}}_0, \tilde{\mathcal{S}}_1, \ldots, \tilde{\mathcal{S}}_T$, with $\tilde{\mathcal{S}}_{t-1} \subset \tilde{\mathcal{S}}_{t}$ for all $t=1, \ldots T$. By definition, $\tilde{\mathcal{S}}_0 = \emptyset$ and $\tilde{\mathcal{S}}_T = \mathcal{T}$. 
We clearly have that $P_\infty ( \tilde{\mathcal{S}}_{t-1} ) \geq P_\infty ( \tilde{\mathcal{S}}_t )$ and $F(\tilde{\mathcal{S}}_t) \geq F(\tilde{\mathcal{S}}_{t-1})$ for all $t=1, \ldots ,  T$.
Note that at stage $t$, the GCC of the network is evaluated by removing all elements in the set $\tilde{\mathcal{S}}_{t-1}$, and only nodes that belong to the current GCC are considered as possible candidates to be added to the structural set $\tilde{\mathcal{S}}_{t}$.

\subsection{Evaluating approximate solutions of the optimal percolation problem}

A possible metric to evaluate the performance of an approximate algorithm to solve the problem of Eq.~(\ref{eq:opt_perc2}) is immediately given by the value of the cost function $F(\tilde{\mathcal{S}}_c)$, with $\tilde{\mathcal{S}}_c$ the approximate solution provided by the algorithm. Low $F(\tilde{\mathcal{S}}_c)$ values indicate a good ability of the algorithm in finding solutions of the optimal percolation model. Specifically, to make the metric comparable across networks and/or variants of the optimal percolation problem, we define the dismantling cost as 
\begin{equation}
q_c = \frac{F(\tilde{\mathcal{S}}_c)}{F(\mathcal{T})}
\; .
 \label{eq:evaluation_a}
\end{equation}
Here, $F(\mathcal{T})$ is the cost associated to the removal of all elements from the graph.

If the algorithm under scrutiny works sequentially by adding to the set of its proposed solution one element at time, then
the quality of the approximate solution of the algorithm can be also quantified by
\begin{equation}
R = 
\frac{1}{F(\mathcal{T})} 
\, \sum_{t = 1}^{T} P_\infty ( \tilde{\mathcal{S}}_t ) \, \left[ F(\tilde{\mathcal{S}}_t) - F(\tilde{\mathcal{S}}_{t-1}) \right]
\; .
    \label{eq:evaluation}
\end{equation}
$R$ is a generalization of the the so-called robustness metric introduced by Schneider {\it et al.}~\cite{schneider2011mitigation}.
By definition $0 \leq R \leq 1$. Low $R$ values are associated to good performance of the dismantling protocol; large $R$ values indicate instead poor performance of the dismantling algorithm. 
The sum appearing in the definition of $R$ is nothing more than the area under the curve $P_\infty(\tilde{\mathcal{S}}_t)$ {\it vs.} $t$. The area is properly rescaled depending on the cost function associated to the dismantling problem. Specifically, the contribution of the element added at the $t$-th stage of the sequential algorithm is proportional to its cost, i.e., $F(\tilde{\mathcal{S}}_t) - F(\tilde{\mathcal{S}}_{t-1})$, and to the
GCC size obtained from
the removal of that set of elements from the network, i.e., $P_\infty ( \tilde{\mathcal{S}}_t )$.
In the standard formulation of the optimal site-percolation problem with unit cost of removal, we recover the original formulation of the metric by Schneider {\it et al.}, i.e., $R = \frac{1}{N} \sum_{t=1}^N P_\infty ( \tilde{\mathcal{S}}_t )$~\cite{schneider2011mitigation}.
For computational reasons, 
in our analysis, we approximate $R$ by summing only the first $T_c$ contributions such that $P_\infty(\tilde{\mathcal{S}}_t) \geq 1 / \sqrt{N}$ for $t = 0, \ldots, T_c$. 
We are basically including only extensive GCCs; this represents a very good approximation for Eq.~(\ref{eq:evaluation}).

\subsection{Algorithms to approximate solutions to the optimal percolation problem}

Many of the algorithms existing in the market are designed to approximate solutions of the optimal site-percolation problem with unit cost of removal. We consider several of them in our analysis. We apply these algorithms without modifications also in the degree-cost version of site percolation problem. Whenever possible, we generalize these algorithms to deal also with the optimal bond-percolation problem with unit cost of removal. We consider three main classes of algorithms, namely (i) baseline, (ii) centrality-based, and (iii) embedding-aided algorithms.

\subsubsection*{{\bf Baseline dismantling algorithms}}

The two algorithms described below represent natural terms of comparisons for generic dismantling algorithms.

{\bf Random percolation (RND).} To generate a baseline solution to the optimal percolation problem, we order the elements of the network randomly. These elements are added sequentially to form the structural sets $\tilde{\mathcal{S}}_t$, for $t=0, \ldots, T$. RND provides a lower bound of performance in the sense that any dismantling algorithm should work at least as good as RND.

{\bf Simulated annealing (SA).}
The algorithm was first introduced in Ref.~\cite{NetDismantling} to deal with optimal site percolation with unit cost. We generalize it to the other variants of the optimal percolation problem. SA is used to find solutions of the problem of Eq.~(\ref{eq:opt_perc2}) only. 
The set $\tilde{\mathcal{S}}_c$ is obtained by first defining a energy-like function, and then applying standard SA optimization to minimize such a function. 
The energy function is defined as $E(\mathcal{S}, \nu) = \nu \times F(\mathcal{S}) + P_\infty(\mathcal{S})$, i.e., the sum 
of the cost $F(\mathcal{S})$ associated to the removal of the set $\mathcal{S}$ and the size of $P_\infty(\mathcal{S})$ of the GCC that the set induces in the network. The two terms compete one against the other, as the goal of the energy minimization is to select a cheap set $\mathcal{S}$ which significantly reduces the size of the GCC. The relative weight of the two terms in the definition of the energy is controlled by the parameter $\nu$, which, depending on the type of dismantling, is chosen in the interval $\nu\in[0.1,1.5]$. 
 This definition is used for all variants of the percolation problem. 
SA provides an upper bound of performance in the sense that we expect other dismantling algorithms to provide solutions less optimal than SA.

\subsubsection*{{\bf Centrality-based dismantling algorithms}}

All dismantling algorithms belonging to this class construct structural sets sequentially, meaning that nodes are ranked according to some specific recipe and added one by one to the structural set. Specifically, if $r_1, r_2, \ldots, r_e, \ldots, r_T$ denote the labels of the ranked elements, then $\tilde{\mathcal{S}}_t = \bigcup_{e =1}^t \{r_e\}$.

{\bf High degree (HD).} For site percolation, we rank nodes in descending order based on the value of their degree centrality, with eventual ties  randomly broken.  For bond percolation, we assign to the edge $(i,j)$ a centrality score $\sigma_{ij} = \frac{k_i \, k_j}{k_i+k_j}$.
Edges are ranked on the basis of their $\sigma$ scores in descending order, and eventual ties are randomly broken. 
Adaptive versions of the above algorithms are obtained by recomputing nodes' degrees only over elements that are not yet part of the structural set. These adaptive versions require a similar computational time as their static counterparts. In our analysis, we use the adaptive version of the algorithm, and refer to it as HDA.

{\bf Collective influence (CI).}
We use also the adaptive version of the so-called collective influence (CI) centrality~\cite{Morone2015}. We use the metric only to approximate solutions of the optimal site-percolation problem. The metric extends HDA. For each node $i$, one first computes the set $\partial \mathcal{B}(i,\ell)$ of all nodes that are at exactly distance $\ell$ from the focal node $i$; CI is then defined as $\sigma_i = (k_i-1) \, \sum_{j \in \partial \mathcal{B}(i,\ell)} (k_j -1)$.  $\ell$ is a tunable parameter. For $\ell = 0$, the metric reduces to HDA. For $\ell =1$, the score reduces to $\sigma_i = (k_i-1) \, \sum_{j} A_{ij} (k_j -1)$. CI  can be computed in a time that scales linearly with the network size~\cite{morone2016collective}. In our tests, we set $\ell = 3$ and $\ell = 4$. Results reported in the paper correspond to the best-performing $\ell$ value.

{\bf Nonbacktracking (NBTC).}
In bond percolation, we rank edges in descending order on the basis of their nonbacktracking centrality (NBTC) scores. Ties are broken at random. The scores are obtained by finding the principal eigenvector $\vec{v}$ of the nonbacktracking matrix of the graph~\cite{hashimoto1989zeta}. The vector contains two components for the edge $(i,j)$, namely $v_{i \to j}$ and $v_{j \to i}$; we associate to the edge $(i,j)$ the score $\sigma_{ij} = \max \{ v_{i \to j}, v_{j \to i} \}$. 
For site percolation, the NBTC of node $i$ is computed as $\sigma_i = \sum_j v_{j \to i}$~\cite{martin2014localization}; nodes are ranked in descending order on the basis of their NBTC, with ties randomly broken. NBTC-based dismantling has been first considered in Ref.~\cite{NetDismantling}. An adaptive version of NBTC may be used too. Our results correspond to the the adaptive version of NBTC.

{\bf Core High Degree (COREHD) and MIN-SUM Decycling.} For site percolation, we use the approach proposed in Ref.~\cite{COREHD} consisting of two main steps. First, we compute the $2$-core of the graph. Then all nodes in the $2$-core are ranked on the basis of their degree centrality, and added to the structural set in descending order. The result of the removal of all nodes in the $2$-core is a tree. The second step of the recipe is a greedy algorithm able to optimally dismantle such a tree~\cite{NetDismantling}.
The idea of dismantling a network by first removing any cycle from it was proposed in Ref.~\cite{NetDismantling}. There, optimal decycling is performed using a MIN-SUM optimization algorithm, consisting in a system of message-passing equations that can be solved in linear time. Details are not included here for sake of brevity. After decycling, the remaining tree is dismantled using the method of Ref.~\cite{NetDismantling}. We use the MIN-SUM algorithm only in the site-percolation variants of the problem.
For bond percolation, we still use COREHD with the only difference that we are allowed to remove links rather than nodes. Namely, we consider only the links within the $2$-core of the network with inclusive degree of its end nodes. By inclusive we mean only nodes that belong to the $2$-core are considered in the computation of the degrees. Then, we assign the score $\sigma_{ij}=\max \{k_i, k_j\}$ to the edge $(i,j)$ and add them to the structural set in desceding order. We adaptively remove links with highest score until the $2$-core disappears from the network. If the size of the GCC is still bigger than a predefined threshold value, we complement COREHD with HDA to dismantle the GCC below the predefined threshold.

{\bf Explosive Percolation (EP).}
For bond percolation, we rely on the EP algorithm proposed in Ref.~\cite{DSouza2015}, here briefly summarized. At the beginning of the algorithm, all edges of the network are considered not active and each node is part of its own component. Edges are activated one by one. The activation of one edge may lead to the merger of two clusters. 
At the $t$-th stage of the algorithm, the score of the edge $(i,j)$ is $\sigma_{ij} = 1 / (c_i \, c_j)$, where $c_i$ is the size of the cluster which node $i$ belongs to. A maximum of $M=1,000$ edges are selected at random among those still not active; the edge with maximum score (ties are randomly broken) is activated, and the score of all other edges is recomputed. The algorithm is iterated until all edges are active. Solutions to the dismantling problem are obtained by reversing the order of activation of the edges in the EP algorithm. 
For site percolation, we rely on a very similar algorithm known in the literature as Explosive Immunization (EI)~\cite{EI}.


\subsubsection*{\bf{Embedding-aided dismantling algorithms}}

We assume that the network is embedded in some geometric space. In the embedding, every node $i$ is mapped to a point $\vec{v}_i$ in the underlying space. The embedding is used to perform iterative bisections of the network. 

For bond percolation, we use the following procedure.
Indicate with $C_z$ and $\mathcal{E}_z$ the 
total number of clusters and the inter-clusters edges 
identified at stage $z$ of the algorithm, respectively.
$T_z$ is the size of the structural set at stage $z$, i.e., $|\tilde{\mathcal{S}}_{T_z} | = T_z$.
The structural set is initialized to $\tilde{\mathcal{S}}_{0} = \tilde{\mathcal{S}}_{T_0} = \emptyset$. Without loss of generality, we assume that at stage $z=1$, the network is  composed of one single cluster $C_1 = 1$. At each stage $z$ of the algorithm, we follow these steps:

\begin{enumerate}
    \item We identify the largest cluster, say $c_{z}$, among the $C_z$ available. We use the already available embedding (or we recalculate the embedding, depending on the specific algorithm) of the nodes in this cluster to find a bipartition. 
    The bipartition of the cluster is obtained considering only elements that do not belong to the set $\tilde{\mathcal{S}}_{T_{z-1}}$. The operation allows us to find two new clusters, thus $C_{z+1} = C_z + 1$ clusters.
    
    \item  We identify all $\mathcal{E}_z$ edges connecting the two parts of cluster $c_z$ determined at step 1. These are edges that stand in-between the two clusters that will originate from $c_z$ but that are not yet part of the structural set, i.e., $e \in \mathcal{E}_z \to e \notin \tilde{\mathcal{S}}_{T_{z-1}}$.
    
    \item We add all edges within $\mathcal{E}_z$ to the structural set in random order. The structural set at this point is $\tilde{\mathcal{S}}_{T_z}$ with size $T_z = \sum_{r=1}^z | \mathcal{E}_r|$. 
    
    \item We increase $z \to z+1$.
    
\end{enumerate}
The algorithm is iterated until all edges are part of the structural set. 

In site percolation, the procedure is analogous. The main difference is that the two clusters that are formed at each iteration should be disconnected by removing nodes rather than edges. 
Please note that finding the minimum number of nodes to be removed in order to disconnect the two clusters is an NP-hard problem known in the literature as minimum vertex cover problem. Here,
we rely on the approximate algorithm 
developed by Ren {\it et al.}~\cite{GNetDismantling}.

{\bf Laplacian embedding (LE).}
This embedding has been considered by Ren {\it et al.} in the context of the site percolation problem, and later generalized by some of the same authors to bond percolation~\cite{GNetDismantling, ren2018underestimated}. 

Nodes are  embedded in a one-dimensional space, where the position of node $i$ is identified by the $i$th component of the eigenvector corresponding to the second smallest eigenvalue of the generalized Laplacian operator $L = D - B$. Here, the $ij$th component of the matrix $B$ is  defined as $B_{ij} = A_{ij} ( c_i + c_j - 1)$; $c_i$ is the cost of removal of node $i$, i.e., $F(\{i\}) = c_i$; $D$ is the diagonal matrix whose $i$th diagonal element is $D_{ii} = \sum_{j} A_{ij}$. The bipartition of the network is obtained by separating nodes on the basis of the sign of their components in the eigenvector. The eigenvector is recomputed at each stage of the dismantling algorithm. 
For bond percolation, the same procedure as above is followed with the only caveat that the embedding of nodes is performed using the standard combinatorial Laplacian~\cite{ren2018underestimated}.

{\bf Hyperbolic embedding (HYP).}
Each node $i$ is mapped to a point $ \vec{v}_i = (r_i, \theta_i)$ in the hyperbolic disk. To perform the embedding, we rely on the so-called Mercator method~\cite{Garc_a_P_rez_2019}. Mercator embeds networks with arbitrary degree distributions via  the maximization of the likelihood function
\[
\mathcal L=\prod_{1 \leq j < i \leq N} p(x_{ij})^{A_{ij}}\left[1-p(x_{ij})\right]^{1-A_{ij}},
\]
where the product goes over all node pairs $ij$ in the network, while $p(x_{ij})$ 
is the Fermi-Dirac connection probability given by 
$p(x_{ij})=\frac{1}{1+e^{(x_{ij}-R)/2G}}$. Here, $x_{ij}=r_i+r_j+2\ln{(\Delta\theta_{ij}/2)}$ is approximately the hyperbolic distance~\cite{Krioukov2010PRE} between nodes $i$ and $j$, $\Delta \theta_{ij}=\pi - | \pi -|\theta_i - \theta_j||$ is the angular (similarity) distance, and $R \sim 2\ln{N}$ is the radius of the hyperbolic disk where all nodes reside. The radial coordinate $r_i$ is related to the observed node
degree $k_i$, as $r_i \sim R-2 \ln{k_i}$ and quantifies node popularity~\cite{Papadopoulos2012Nature}.
The value of the temperature parameter $G$  for a given network is also inferred by Mercator.
The maximization of the likelihood function leverages the Laplacian Eigenmaps approach of Ref.~\cite{Muscoloni2017NatComm}.
Hyperbolic coordinates are estimated on the entire network topology. At each stage of the dismantling algorithm, a bipartition is obtained by cutting in half the slice of the hyperbolic disk of the cluster under consideration. 

{\bf Node2vec embedding (N2V).}
Node2vec~\cite{grover2016node2vec} is a network embedding algorithm that builds on the word2vec algorithm~\cite{mikolov2013efficient} by taking the following analogy: nodes in the network are considered as words; a sequence of nodes explored during a biased random walk is considered as a sentence.
Nodes are embedded in the $d$-dimensional Euclidean space. The embedding is dependent on various hyperparameters. We fix the number of walks per node to $10$,  the number of iterations to $10$, and the parameters that bias the random walk toward a breadth-first or depth-first walk both equal to $1$.
Results of some tests reported in Fig.~S12 indicate that optimal dismantling is achieved for large values of the embedding dimension $d$, and medium values of the walk length $l$. We therefore fix 
$d = 2,048$ and  $l=32$ in our analysis. 

The bipartition is obtained using a $k$-means algorithm with $k=2$ clusters~\cite{lloyd1982least}. Clusters are created on the basis of the Euclidean distance between nodes in the space. We compute the embedding only once, and then use the same map in all stages of the dismantling algorithm.

{\bf Algorithmic complexity.}
The complexity of an embedding-aided dismantling algorithm is approximately $N \log_2 N$ on sparse networks. This can be understood by thinking the iterative procedure
as equivalent to the generation of a rooted binary tree. The root of the tree is the input network. Intermediate nodes are the clusters obtained during the iterative dismantling algorithm. Leaves are individual nodes. Roots and intermediate nodes have two offsprings corresponding to the split of a cluster in two smaller clusters. The height of such as tree is $H = \log_2 N$. At each level $h$ of the tree there $C_h$ clusters composed of a number of elements proportional to $N / C_h$.
Finding the bipartition of a cluster and determining the inter-cluster edges require a time that grows proportionally to the cluster size, 
therefore each level of the tree is processed in a time that grows as $N$.
The above computation of the complexity assumes that embedding and bisecting a cluster of nodes require a time scaling at maximum with its size. This is true for both the Laplacian and Node2vec embeddings. The computational time required to embed a network in hyperbolic space with the Mercator algorithm scales instead as $N^2$. The quadratic scaling dominates the time complexity of the dismantling algorithm based on hyperbolic embedding.


\subsubsection*{{\bf Greedy post-processing technique}}
\label{GR}

Approximate solutions of the various algorithms are further refined with a simple, but effective greedy post-processing strategy. The general principle is to remove from a potentially spurious structural set all elements that are not necessary to keep the GCC of the network below a certain predetermined value. The strategy is useful to reduce the size of the structural set, and thus obtain a better solution for the problem of Eq.~(\ref{eq:opt_perc2}). The strategy for site percolation was introduced in Refs.~\cite{NetDismantling, Morone2015}. Here, we extend it to bond percolation with unit cost and site percolation with degree cost.


\subsection{Networks}

\subsubsection*{\bf{Real networks}}

We consider a corpus of $50$ real-world networks. Networks have size ranging from $N = 309$ to $N = 62561$. The upper bound on the network size is due to computational reasons, as some of the dismantling algorithms considered in our analysis do not scale well with the system size. Details on the various networks are reported in the SM.

\subsubsection*{\bf{Synthetic networks}}

{\bf $\mathbb{H}^2$ model.} We create 
instances of the $\mathbb{H}^2$ model~\cite{Krioukov2010PRE, aldecoa2015hyperbolic}
with $N=2^{14}$ nodes, degree exponents $\gamma \in \{2.2, 2.6, 3.5\}$, average degree $\langle k \rangle \approx 6$, and values of the temperature parameter $G \in \{0.1, 0.2, \dots, 0.9\}$. 
The parameter $\gamma$ controls the heterogeneity of the degree distribution, as $P(k) \sim k^{-\gamma}$ for networks generated according to this model. The temperature parameter $G$ controls the strength of correlation between network topology and imposed embedding, with low values of $G$ favoring connections between pairs of nodes at small hyperbolic distance.

{\bf Lancichinetti-Fortunato-Radicchi (LFR) model.}
We create networks according to the LFR model~\cite{lancichinetti2008benchmark} with $N=2^{14}$ nodes and degree exponent $\gamma \in \{2.2, 2.6, 3.5\}$, average degree $\langle k \rangle = 6$, maximum degree $k_{max}=\sqrt N$. We use values of the mixing parameter $\mu \in \{0.05, 0.1, \dots, 0.5\}$. Communities are distributed randomly with size distribution $P(s) \sim s^{-1}$ with $\sqrt N$ and $5 \times \sqrt N$ chosen as minimum and maximum size of a community, respectively. Also for the LFR model, the parameter $\gamma$ controls the heterogeneity of the degree distribution, i.e., $P(k) \sim k^{-\gamma}$. The mixing parameter $\mu$ controls the strength of correlation between network topology and imposed embedding, as low $\mu$ values favor connections between pairs of nodes belonging to the same pre-imposed communities.

\section*{Acknowledgments}
F.P. acknowledges support by the TV-HGGs project (OPPORTUNITY/0916/ERC-CoG/0003), co-funded by the European Regional Development Fund and the Republic of Cyprus through the Research and Innovation Foundation. F.R. acknowledges support by the Air Force Office of Scientific Research (FA9550-21-1-0446) and by the Army Research Office (W911NF-21-1-0194). The funders had no role in study design, data collection and analysis, decision to publish, or  any opinions, findings, and conclusions or recommendations expressed in the manuscript.


%

\end{document}